\documentclass{article}

\usepackage{arxiv}

\usepackage[utf8]{inputenc} 
\usepackage[T1]{fontenc}    
\usepackage{url}            
\usepackage{amsfonts}       
\usepackage{nicefrac}       
\usepackage{microtype}      

\usepackage{amsmath,amssymb,amsfonts,amsthm,nccmath}
\usepackage{algorithm,algpseudocode}
\usepackage{graphicx}
\usepackage{textcomp}
\usepackage{xcolor}
\usepackage{array, booktabs, boldline, threeparttable}
\usepackage{enumitem}
\usepackage{float}

\newtheorem{definition} {Definition}  
\newtheorem{example} {Example}

\algnewcommand{\IIf}[1]{\State\algorithmicif\ #1\ \algorithmicthen}
\algnewcommand{\EndIIf}{\unskip\ \algorithmicend\ \algorithmicif}

\title{BioMETA: A multiple specification parameter estimation system for stochastic biochemical models}

\author{
  Arfeen Khalid \\
  Department of Computer Science\\
  University of Central Florida\\
  Orlando, FL 32816 \\
  \texttt{akhalid@cs.ucf.edu} \\
}

\begin{document}
\maketitle

\begin{abstract}

The inherent behavioral variability exhibited by stochastic biochemical systems makes it a challenging task for human experts to manually analyze them. Computational modeling of such systems helps in investigating and predicting the behaviors of the underlying biochemical processes but at the same time introduces the presence of several unknown parameters. A key challenge faced in this scenario is to determine the values of these unknown parameters against known behavioral specifications. The solutions that have been presented so far estimate the parameters of a given model against a single specification whereas a correct model is expected to satisfy all the behavioral specifications when instantiated with a single set of parameter values. We present a new method, BioMETA, to address this problem such that a single set of parameter values causes a parameterized stochastic biochemical model to satisfy all the given probabilistic temporal logic behavioral specifications simultaneously. Our method is based on combining a multiple hypothesis testing based statistical model checking technique with simulated annealing search to look for a single set of parameter values so that the given parameterized model satisfies multiple probabilistic behavioral specifications. We study two stochastic rule-based models of biochemical receptors, namely, Fc$\epsilon$RI and T-cell as our benchmarks to evaluate the usefulness of the presented method. Our experimental results successfully estimate $26$ parameters of Fc$\epsilon$RI and $29$ parameters of T-cell receptor model against three probabilistic temporal logic behavioral specifications each.



\end{abstract}


\section{Introduction}
Stochastic rule-based models serve as a natural and compact representation for biochemical reactions. Biochemical modeling languages \cite{faeder2009rule,blinov2004bionetgen} are used to succinctly describe biochemical systems, such as cell signaling pathways. The  Gillespie stochastic simulation algorithm \cite{gillespie1977exact} and its variants \cite{gillespie2009refining} are then employed to predict the behavior of the biochemical system modeled by the stochastic rule-based model.

However, it is often not feasible to create a complete stochastic rule-based model from first principles. Instead, our knowledge of the biochemical system is used to obtain the set of chemical reactions or the core structure of the stochastic rule-based model. The lack of knowledge about the rate constants of biochemical reactions is readily  modeled as unknown parameters in stochastic rule-based models. 

A primary challenge in the use of such a parameterized stochastic rule-based model for predicting the behavior of a biological system is the determination of the parameters of the model from multiple experimental observations. Traditionally, parameter values of a stochastic model have been estimated against quantitative time series data. A few relatively recent efforts \cite{calzone2006machine, hussain2015automated} have focused on estimating the parameters of a biological model against qualitative stochastic experimental observations encoded as probabilistic temporal logic formulas. 

However, the focus of these efforts has been on discovering parameter values of a paramaterized stochastic biological model from a single probabilistic temporal logic specification. In practice, a biological model must satisfy multiple experimental observations made on the biological system being modeled. Hence, it is important to estimate a single set of parameter values that causes a parameterized stochastic model to satisfy multiple probabilistic temporal logic specifications simultaneously.

In this study, we make the following contributions towards parameter estimation of stochastic biochemical models:

\begin{enumerate}[label=(\roman*)] 
\item We design a new approach that utilizes a \emph{quantitative tightness metric} to estimate a single set of parameter values for a parameterized  stochastic biochemical model such that the model satisfies all the given probabilistic temporal logic behavioral specifications \emph{simultaneously}. Our approach employs multiple hypothesis testing based statistical model checking to  evaluate stochastic models against multiple specifications. It further uses simulated annealing search to estimate the unknown parameters present in the model.
\item Our experimental results demonstrate that our method BioMETA is able to estimate $26$ parameters of the Fc$\epsilon$RI model and $29$ parameters of the T-cell receptor model satisfying three temporal logic properties each. \end{enumerate}

\section{Background}
In the following subsections, we first describe rule-based modeling using software tool BioNetGen \cite{blinov2004bionetgen}. We then formally define Signal Temporal Logic (STL) \cite{donze2013efficient} that is used to encode behavioral specifications as temporal logic formulas. We also describe TeLEx \cite{jha2017telex} that computes the quantitative tightness metric describing how well the given model satisfies a behavioral specification.

\subsection{Rule-based modeling}
Rule-based modeling is a relatively new formalism that allows to model biochemical systems by representing molecules as structured objects and molecular interactions as rules \cite{faeder2009rule,chylek2014rule,xu2011rulebender,smith2012rulebender}. The interactions can be of various types including associations, dissociations, modifications to the internal state of a molecule as well as the production or consumption of molecular species \cite{liu2016parameter}. In other words, a rule specifies how the states of reactants are modified to generate products in a biochemical reaction. Molecular interactions can result in a large number (hundreds to thousands) of possible molecular species. Conventional approaches, involving translating the model into ODEs or CTMCs, are unable to handle such combinatorial complexity. Rule-based modeling addresses this problem by expressing models with a high degree of modularity and avoid the explicit enumeration of all possible molecular species or all the states of a system; hence, providing a succinct representation of the model \cite{faeder2009rule}. The rules are then simulated to generate a reaction network comprised of all chemically distinct species and reactions \cite{blinov2004bionetgen}. Therefore, a rule-based model is a compact and generalized representation of conventional biochemical models.

\subsection{BioNetGen software}
Biological Network Generator (BioNetGen) is an open source software tool that can be used to construct, visualize, simulate and analyze rule-based models. The tool is based on a formal language, known as BioNetGen language (BNGL), for specifying molecules and rules to model biological systems. A BNGL model file is composed of six primary blocks \cite{blinovrule}, namely, \textit{(i) parameters} that include rate constants and values for initial concentrations of species and are responsible for governing the dynamics of the system, \textit{(ii) molecule types} defining molecules, including components and allowed component states, \textit{(iii) seed species} that describe the initial state of the system, \textit{(iv) observables} that output functions of concentrations of species having particular attributes, \textit{(v) reaction rules} describing how molecules interact with each other and \textit{(vi) actions} that provide various methods for generating and simulating the network. In our experiments, we use the stochastic simulation algorithm (SSA) implemented in the software to simulate our biochemical models. Each model simulation results in a time-stamped simulation trace capturing the behavior of the model when instantiated with a particular set of parameter values.

\subsection{Signal Temporal Logic and TeLEx}
It is required that the acceptable behaviors of a model are represented as temporal logic formulas so that computational methods can be applied to verify whether a model satisfies a behavior or not. In this study, we use Signal Temporal Logic (STL) to encode the known behaviors of a stochastic biochemical model. It is a linear time temporal logic used to represent continuous time behaviors of a model \cite{raman2015reactive,donze2013efficient}. We use a time bounded variant of STL where all temporal operators are associated with a lower and upper time-bound. The logical operators in STL formulas consist of $\land$ (and), $\lor$ (or), $\neg$ (negation), and bounded time operators consist of \textbf{G} (global), \textbf{F} (future), and \textbf{U} (until).

\begin{definition} [Signal Temporal Logic] A Signal Temporal Logic formula representing a model's known behavior is defined recursively by the following grammar:
\end{definition}


\begin{alignat*}{3}
	\phi := \ \mu \ | \ \neg \mu \ | \ \phi_1 \land \ \phi_2 \ | \ \phi_1 \lor \ \phi_2 \ |
	 \ \text{G}_{[t_1, t_2]} \ \phi \ | \ \text{F}_{[t_1, t_2]} \ \phi \ | \ \phi_1 \ \text{U}_{[t_1, t_2]} \ \phi_2
\end{alignat*}

\noindent where  $\phi$,  $\phi_1$ and $\phi_2$ are STL formulas, 0 $\leqslant$ $t_1$ $<$ $t_2$ $<$ $\infty$ and $\mu$ is a predicate whose value is determined by the sign of a function of an underlying \textit{signal} \textbf{x} $\in \mathbb{R}$, i.e., $\mu$ $\equiv$ $\mu(\textbf{x})$ $>$ 0 \cite{raman2015reactive,jha2017telex}. In this study, signal \textbf{x} is termed as a \emph{simulation trace} or \emph{simulation trajectory} and is obtained by simulating a given model in BioNetGen software. The validity of a formula $\phi$ with respect to signal \textbf{x} at time $t$ is defined inductively as follows:

\begin{alignat*}{4}
	&(\textbf{x}, t) \vDash \mu  && \Leftrightarrow \mu(x_t) > \text{0} \\
	&(\textbf{x}, t) \vDash \neg \mu && \Leftrightarrow \neg((\textbf{x}, t) \vDash \mu) \\
	&(\textbf{x}, t) \vDash \phi_1 \land \phi_2\ && \Leftrightarrow (\textbf{x}, t) \vDash \phi_1 \land (\textbf{x}, t) \vDash \phi_2 \\
	&(\textbf{x}, t) \vDash \phi_1 \lor \phi_2 && \Leftrightarrow (\textbf{x}, t) \vDash \phi_1 \lor (\textbf{x}, t) \vDash \phi_2 \\
	&(\textbf{x}, t) \vDash \text{G}_{[t_1, t_2]} \phi &&  \Leftrightarrow \forall t^\prime \in [t+t_1, t+t_2] \text{ s.t.} (\textbf{x}, t^\prime) \vDash \phi \\
	&(\textbf{x}, t) \vDash \text{F}_{[t_1, t_2]} \phi && \Leftrightarrow \exists t^\prime \in [t+t_1, t+t_2] \text{ s.t.} (\textbf{x}, t^\prime) \vDash \phi \\
	&(\textbf{x}, t) \vDash \phi_1 \text{U}_{[t_1, t_2]} \phi_2 && \Leftrightarrow \exists t^\prime \in [t+t_1, t+t_2] \text{ s.t.} (\textbf{x}, t^\prime) \vDash \phi_2 \\
	&&& \quad \quad \land \forall t^{\prime\prime} \in [t, t^\prime], (\textbf{x}, t^{\prime\prime}) \vDash \phi_1
\end{alignat*}

A signal \textbf{x} satisfies $\phi$, denoted by \textbf{x} $\vDash$ $\phi$, if (\textbf{x},0) $\vDash$ $\phi$. 
Informally, \textbf{x} $\vDash$ $\text{G}_{[t_1, t_2]} \phi$ if $\phi$ holds at every time
step between $t_1$ and $t_2$, and \textbf{x} $\vDash$ $\text{F}_{[t_1, t_2]} \phi$ if $\phi$ holds at some time step between $t_1$
and $t_2$. Also, \textbf{x} $\vDash$ $\phi_1 \text{U}_{[t_1, t_2]} \phi_2$ if $\phi_1$ holds at every
time step before $\phi_2$ holds, and $\phi_2$ holds at some time step between
$t_1$ and $t_2$. 

Due to the stochastic nature of the given biochemical models, their behavioral specifications are also often probabilistic in nature. Therefore, we use a probabilistic variant of Signal Temporal Logic to encode model behaviors. Probabilistic STL can be further explained with the help of following example:

\begin{example}
Consider the following probabilistic Signal Temporal Logic formula:
\end{example}

\begin{center}
    $P_{\geq 0.85}(\textbf{G}_{[0,100]}((GProtein > 6000) \hspace{1mm} \land$ 
      $\textbf{F}_{[100,200]}(GProtein < 6000)))$
\end{center}


\textit{It says that G protein should always have a high value i.e. greater than 6000 units during the first 100 (0-100) time units and should fall below 6000 at some point in time during the next 100 (100-200) time units with a probability of at least 0.85.}\\

To verify the model against a given STL property we use the algorithm implemented by a software framework Temporal Logic Extractor (TeLEx) \cite{jha2017telex}. Given a time-stamped simulation trace and a STL behavioral specification formula, TeLEx quantifies the degree of \emph{model satisfiability} by computing a tightness metric that describes how well a model satisfies the specification. TeLEx uses smooth functions, such as sigmoid and exponentials, to compute tight-satisfiability of STL formulas. A \emph{successful simulation trajectory} of a model refers to such a trajectory that satisfies a given STL formula. Whereas an \emph{unsuccessful simulation trajectory} refers to a trajectory that does not satisfy the given STL formula. TeLEx returns a positive value in case the STL formula is verified successfully; the larger the value the better it satisfies the behavior. If the model is not able to satisfy the STL formula, the algorithm returns a negative value describing how far off is the model from satisfying the specification; for further details and examples see \cite{jha2017telex}.

\section{Related Work}
One of the crucial steps in the process of parameter estimation is to check the model against behavioral temporal logic specifications. Statistical Model Checking (SMC) \cite{legay2010statistical} is a popular method among many well-studied model checking techniques available in the literature. A detailed survey \cite{zuliani2015statistical} on various SMC techniques indicates that Statistical Model Checking based on Wald's Sequential Probability Ratio Test (SPRT) \cite{wald1945sequential} is widely studied \cite{palaniappan2013statistical,hussain2014parameter,ramanathan2015parallelized} and used in practice. Another variant of SMC \cite{jha2009bayesian,hussain2015automated} uses Bayesian Sequential Hypothesis Testing and improves performance by incorporating prior knowledge about the model being verified. Mancini et al. \cite{mancini2015computing} propose a parallel SMC algorithm to yield model parameters of biological systems represented as Ordinary Differential Equations (ODEs). Their approach follows a master-slave architecture where a single master implements the SMC algorithm and assigns the numerical integration of the given ODE system to slaves.

Satisfiability Modulo Theory (SMT) has also been used to perform model checking of biological models \cite{madsen2015biopsy}. It works by first extracting a collection of ODEs from a given model and then formulating these ODEs along with time-series data into a collection of SMT problems. Another approach \cite{calzone2006machine} performs model checking based on an open source symbolic model checker, NuSMV \cite{cimatti2002nusmv}, and formulates a given model using Binary Decision Diagrams (BDDs). These BDDs provide a compact way of representing boolean functions which in turn represent the different states of the given model.
 
A recent study \cite{liu2016parameter} presents a SMC based parameter estimation framework for rule-based models formulated in BioNetGen. It verifies experimental data as well as qualitative properties of the given model. Wang et al. \cite{wang2016formal} extends the rule-based BioNetGen language to enable the specification of interactions among more than one cell. They employ SMC in order to analyze system properties and obtain interesting insights into the development of
novel therapeutic strategies for pancreatic cancer. Techniques based on applying convolutional neural networks \cite{zhou2017automated} in order to learn temporal formulas have also been recently proposed. 

However, each of these methods is limited to verifying the model with a binary (yes/no) outcome whereas we compute a quantitative tightness metric describing how well the given model satisfies a behavioral specification. Further, these methods are limited to verifying only a single temporal logic property of the model at one time whereas BioMETA uses a sequential multiple hypothesis testing technique to validate the model against all given specifications simultaneously.

\subsection{Approaches using a quantitative measure to check model satisfiability}

Several attempts have been made to estimate parameters of stochastic systems using a quantitative measure of how well a model satisfies a known temporal logic behavioral specification.

Rizk et al. \cite{rizk2008continuous} define a continuous degree of satisfaction of a temporal logic property which is then used as a fitness function in order to find kinetic parameters of a biochemical model. However, their continuous degree of satisfaction is limited to providing a quantitative measurement only in case of dissatisfaction and remains zero in case the model satisfies a temporal logic property. Whereas we compute a tightness metric that provides a quantitative measure irrespective of whether or not the model satisfies a system property; thus, giving a more meaningful interpretation of model verification.

Another technique \cite{bartocci2013robustness} using Gaussian Process Upper Confidence Bound (GP-UCB) computes a distribution of a quantitative satisfaction function specifying the degree of satisfaction of a property by the model. An average of this distribution is later used to guide the parameter search. However, this approach is limited to verifying only one temporal logic property at one time as opposed to BioMETA's ability to verify all the properties simultaneously.

A global exploration of the parameter space of stochastic biochemical systems is performed using probabilistic model checking \cite{brim2013exploring,vcevska2017precise,vcevska2016prism}. For each parameter point, they compute approximate upper and lower bounds of a landscape function that returns a quantitative value. This value is based on the probability that the model satisfies a given CSL (continuous stochastic logic) formula. Another software framework uses the Bayesian formalism for parameter estimation and model selection \cite{liepe2014framework}. Euclidean distance (sum of squares) between the observed data and a simulated trajectory is computed; a parameter point is accepted if this distance is less than a threshold value.

To the best of our knowledge, this is the first time that a quantitative tightness metric is used to check model satisfiability against multiple probabilistic temporal logic behavioral specifications simultaneously; hence, generating a single set of parameter values.

\section{Our Approach}
We are interested in finding a set of parameter values $\rho_0 \in \mathbb{R}^n$ so that the model $\mathcal{M}$, when instantiated with set $\rho_0$, satisfies each specification $\phi_{i}$ with probability greater than or equal to its corresponding required probability $r_{i}$. We now formally define our problem.

\begin{definition}
[Multiple specification parameter estimation problem] Given a parameterized stochastic rule-based model $\mathcal{M}$($\rho$) with parameter set $\rho \in \mathbb{R}^n$, a set of  desired specifications $\pmb{\phi}$ = \{$\phi_1,..., \phi_k$\} in Signal Temporal Logic, and a corresponding set of required probabilities $\pmb{r}$ =  \{$r_1,..., r_k$\} where $r_i \in$ [0,1], find a set of parameter values $\rho_0$ such that the following holds:
$$\mathcal{M}(\rho_0) \vDash P_{\geq r_1}(\phi_1) \land P_{\geq r_2}(\phi_2) \land \dots \land P_{\geq r_k}(\phi_k)$$
\end{definition}

\begin{figure}[h!]
	\centering
	\includegraphics[width=0.65\columnwidth]{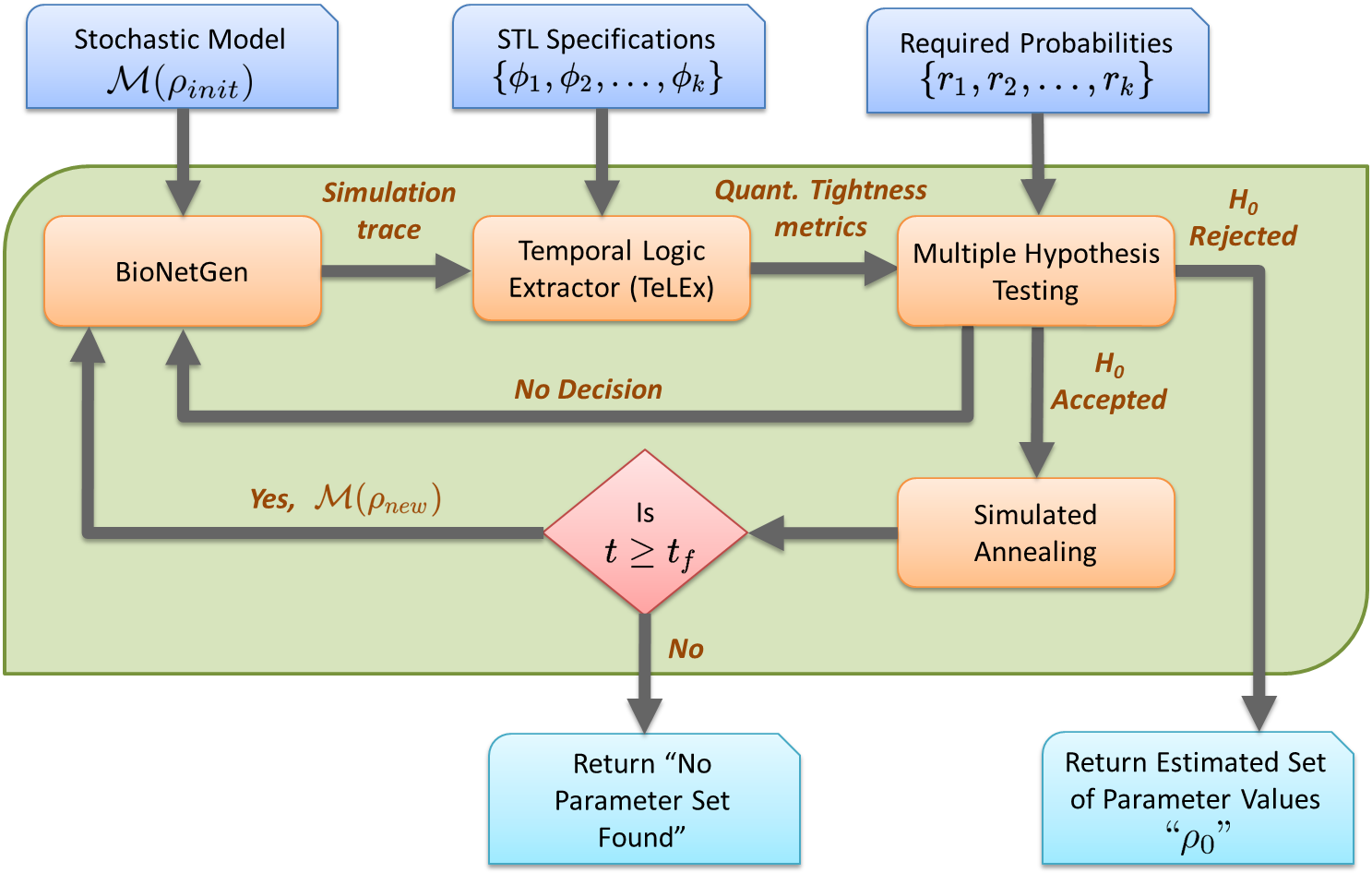}
    \caption{Our proposed method: BioMETA.
    A Multiple specification parameter estimation system using multiple hypothesis testing based statistical model checking.}
    \label{figure1}
\end{figure}

Our solution to the problem of estimating parameters from multiple specifications is illustrated in Figure \ref{figure1}. Given an initial set of parameter values $\rho_{init}$, we first simulate the given model $\mathcal{M}$($\rho_{init}$) using BioNetGen to generate a time-stamped simulation trace. The simulation trace along with the given set of $k$ STL specifications $\pmb{\phi}$ = \{$\phi_1,..., \phi_k$\} is then fed to TeLEx which returns $k$ quantitative tightness metrics describing the distance between the model $\mathcal{M}$($\rho_{init}$) and each of the $k$ given specifications. TeLEx quantitative tightness metrics and the set of required probabilities $\pmb{r}$ =  \{$r_1,..., r_k$\} are then passed onto multiple hypothesis testing (MHT) that decides whether the model instantiated with $\rho_{init}$ satisfies the given probabilistic STL specifications or not. The hypothesis test continues on drawing samples, i.e. continues on generating model simulations, until it has made a decision. If the test declares that $\mathcal{M}$($\rho_{init}$) satisfies all the given specifications with greater than or equal to their corresponding required probabilities, then $\rho_{init}$ is returned as the estimated set of parameter values. Otherwise a search algorithm, guided by the mean of quantitative tightness metrics, explores the parameter space of the given model in order to find a new set of parameter values and repeats the above process.

We explain the approach implemented in this study by dividing it into two phases. The first phase employs multiple hypothesis testing (MHT) based statistical model checking to check a given model against multiple STL behavioral specifications. The second phase of our method implements a simulated annealing based search algorithm to explore model's parameter space and finds a single set of parameter values which satisfies all the given probabilistic behavioral specifications simultaneously.

\subsection{Multiple hypothesis testing based model checking}

As the problem (Definition 2) states, given a set $\pmb{\phi}$ of $k$ STL behavioral specifications we are required to generate a single set of parameter values such that the model satisfies each specification $\phi_i$ with probability at least $r_i$. In order to achieve this, the given model is first simulated using BioNetGen and the simulation trace is verified using TeLEx against each given specification. Essentially, TeLEx verification function is called $k$ times and each call returns a quantitative tightness metric against the respective specification $\phi_i$. 

Due to the stochastic nature of the given model, model simulations with the same parameter values show varying behaviors which results in varying TeLEx tightness metrics. Therefore a multiple hypothesis testing technique, proposed by Bartroff et al. \cite{bartroff2014sequential}, is employed that generates several model simulations before deciding model satisfiability. The process of generating multiple simulations results in forming a separate distribution of TeLEx tightness metrics against each specification. Hence, $k$ distributions are generated where each distribution corresponds to the tightness metrics computed by TeLEx against each of the $k$ STL specifications.  

We define the mean of each distribution $i$ as $\theta^{(i)}$. In order to decide whether the given model satisfies a specification, we aim to test the hypothesis that the mean $\theta^{(i)}$ of each distribution is less than a threshold value $\theta'^{(i)}$ \cite[Section 5.4]{wald1945sequential}. If the mean $\theta^{(i)}$ of distribution $i$ is greater than its corresponding threshold $\theta'^{(i)}$, the model satisfies $\phi_i$ since a greater mean of TeLEx tightness metrics indicates better model satisfiability. However if $\theta^{(i)} < \theta'^{(i)}$, it is considered that the model does not satisfy the corresponding specification $\phi_i$. 

Instead of testing the hypothesis against a single value $\theta'^{(i)}$, the problem is relaxed by introducing two thresholds $\theta_0^{(i)}$ and $\theta_1^{(i)}$ such that $\theta_0^{(i)} < \theta'^{(i)} < \theta_1^{(i)}$. We vary the two thresholds using a common factor, $\delta$ i.e. $\theta_0^{(i)} = \theta'^{(i)} - \delta$ and $\theta_1^{(i)} = \theta'^{(i)} + \delta$ where $0 < \delta < 1$ \cite{younes2006statistical}. The region $(\theta_0^{(i)}, \theta_1^{(i)})$ is defined as the \textit{indifference region}. As the indifference region becomes smaller, we reduce the risk of making a wrong decision but this comes at the cost of drawing more samples in order for the hypothesis test to make a decision. Hence for each distribution $i$, we test the following hypothesis:

\begin{ceqn} \label{}
    \begin{align} 
        \begin{split}
            H_0^{(i)}: \theta^{(i)} = \theta_0^{(i)} \\
            H_1^{(i)}: \theta^{(i)} = \theta_1^{(i)}
        \end{split}
    \end{align}
\end{ceqn}

\noindent with type I (false negative) and type II (false positive) family-wise error rate (FWER) bounds as $\alpha$ and $\beta$ respectively. Suppose $x_j^{(i)}$ is the tightness metric returned by TeLEx after verifying $j^{th}$ simulation trace against $i^{th}$ STL specification $\phi_i$. Then for the $n^{th}$ model simulation, the hypothesis test calculates the following log-likelihood ratio against each specification $\phi_i$:

\begin{ceqn} \label{eq_4}
    \begin{align}
        Z_i = \log \frac{e^{-(1/2\sigma^{2}_i) \sum_{j=1}^n (x_j^{(i)} - \theta_1^{(i)})^2}}
        {e^{-(1/2\sigma^{2}_i) \sum_{j=1}^n (x_j^{(i)} - \theta_0^{(i)})^2}}
    \end{align}
\end{ceqn}

The test samples sequentially until $Z_i \leq A_i$ or $Z_i \geq B_i$ where $A_i$ and $B_i$ are the stopping boundaries and are defined as functions of FWER bounds $\alpha$ and $\beta$. For each distribution $i$, the stopping boundaries are defined as:

\begin{ceqn} \label{eq_2}
    \begin{align}
        \begin{split}
            A_i = log\Bigg( \frac{\beta}{(1-\alpha_i)(k - i + 1)} \Bigg), \quad
            B_i = log\Bigg( \frac{(1-\beta_i)(k - i + 1)}{\alpha} \Bigg)
        \end{split}
    \end{align}
\end{ceqn}

\noindent where,
\begin{ceqn} \label{eq_3}
    \begin{align}
        \begin{split}
            \alpha_i = \frac{(k - i + 1 - \beta)\alpha}{(k - i + 1)(k - \beta)}, \quad
            \beta_i = \frac{(k - i + 1 - \alpha)\beta}{(k - i + 1)(k - \alpha)}
        \end{split}
    \end{align}
\end{ceqn}

Note that the hypothesis test continues on taking samples until each $Z_i$ crosses one of its corresponding stopping boundaries $A_i$ or $B_i$.

Rejection of the null hypothesis $H_0^{(i)}$ indicates that the mean of the distribution $i$ is greater than threshold $\theta'^{(i)}$ and model satisfies the corresponding behavioral specification $\phi_i$ whereas, acceptance of a null hypothesis $H_0^{(i)}$ means that the model with its current set of parameter values is not able to satisfy $\phi_i$. 
In case all $H_0^{(i)}$ are rejected, BioMETA further checks whether each specification $\phi_i$ is satisfied with probability at least $r_i$. If yes, BioMETA returns the current set of parameter values as the solution otherwise, it explores other parameter values using our search algorithm. We describe the behavior of the search algorithm in the following subsection.

\subsection{Parameter search algorithm}

To explore the parameter space of the given model, we employ an iterative search algorithm known as simulated annealing. It is a probabilistic technique for approximating the global optimum of a given function. The algorithm (refer to Algorithm \ref{algorithm1}) starts the search at a high temperature value, $t_i$, provided as input (line 1). It then generates a random set of parameter values, $\rho_{init}$, and checks model satisfiability against given specifications using our multiple hypothesis testing (MHT) based statistical model checking approach (described in the previous subsection) (lines 2-3). $\rho_{init}$ is returned as the estimated set of parameter values in case the hypothesis test returns ``accept" which indicates that the model $\mathcal{M}$($\rho_{init}$) satisfies all given specifications (line 4). If the hypothesis test rejects $\rho_{init}$, the algorithm generates a new neighboring set of parameter values, $\rho_{new}$, by slightly perturbing the previous parameter set, and again checks model satisfiability with $\rho_{new}$. Every time a parameter set is rejected, the algorithm starts a new iteration by finding a neighboring parameter set and repeating the model checking step (lines 7-21). 

The objective function that drives the search algorithm is defined as the \textit{mean of unsuccessful simulations} exhibiting negative TeLEx tightness metrics. For each new iteration, mean ($\mu_{new}$) of the current iteration is compared with mean ($\mu_{old}$) of the previous iteration and the parameter set that is maximizing the mean function is accepted. However, in order to avoid local optima the algorithm also sometimes, with a very small probability, accepts bad solutions i.e. accepts parameter set with smaller means (lines 14-18). As the current temperature $t$ goes down, based on a cooling factor $c$ (line 20), the probability of accepting bad solutions also decreases in order for the algorithm to achieve stability. Once it finds a set of parameter values for which the model successfully satisfies all specifications with their respective required probabilities, it stops and returns that parameter set. The algorithm continues until it finds the required set of parameter values or the given temperature cools down to a predefined value, $t_f$. In the later case, the algorithm returns that it was unable to find the required parameter set (line 22). A python based implementation of Algorithm 1 can be found at \url{https://github.com/arfeenkhalid/parameter-synth}.

\begin{algorithm}
\caption{BioMETA Algorithm: Multiple specification parameter estimation system.}
\label{algorithm1}
\begin{algorithmic}[1]

\Require~~
	\Statex $\mathcal{M}$($\rho$) \hspace*{14.5mm} Parameterized stochastic model
	\Statex \{$\phi_1,\phi_2,..., \phi_k$\} \hspace*{1mm} STL specifications
	\Statex \{$r_1,r_2,..., r_k$\} \hspace*{2.5mm} required probabilities
	\Statex $t_i$ \hspace*{20.5mm} initial temperature
	\Statex $t_f$ \hspace*{20mm} final temperature
	\Statex $c$ \hspace*{21.5mm} cooling rate

\Ensure~~
	\Statex $\rho_0$ such that $\mathcal{M}(\rho_0)$ $\vDash$ $P_{\geq r_1}(\phi_1) \land P_{\geq r_2}(\phi_2) \land .... \land P_{\geq r_k}(\phi_k)$ 
\item[]
\State t $\leftarrow$ $t_i$
\State ${\rho_{init} \leftarrow \textit{rand(\hspace{0.7mm})}}$ 
\State result, $\mu_{init}$ $\leftarrow$ \Call{MHT}{$\mathcal{M}$($\rho_{init}$), \{$\phi_1,.., \phi_k$\}, \{$r_1,.., r_k$\}}
\IIf{result = ``acc"} \Return $\rho_{init}$ \EndIIf 
\State $\rho_{new}$ $\leftarrow$ $\rho_{init}$
\State $\mu_{new}$ $\leftarrow$ $\mu_{init}$
\While{t $\geqslant$ $t_f$}
	\State $\rho_{old}$ $\leftarrow$ $\rho_{new}$
    \State $\mu_{old}$ $\leftarrow$ $\mu_{new}$
	\State $\rho_{new}$ $\leftarrow$ \Call{FindANeighbour}{$\rho_{old}$}
    \State result, $\mu_{new}$ $\leftarrow$ \Call{MHT}{$\mathcal{M}$($\rho_{new}$), \{$\phi_1,.., \phi_k$\}, \{$r_1,.., r_k$\}}
    \If{result = ``acc"}
		\Return $\rho_{new}$
    \Else
    	\If{$\mu_{new}$ $\leqslant$ $\mu_{old}$}
        	\If{\textit{rand}(0, 1) $>$ $e^{(-(\mu_{new} - \mu_{old}  )/t)}$}
            	\State  $\rho_{new}$ $\leftarrow$ $\rho_{old}$
        	\EndIf
        \EndIf
	\EndIf
    \State t $\leftarrow$ $c$ $\ast$ t
\EndWhile \\
\Return ``No parameter set found!''
\end{algorithmic}
\end{algorithm}

\section{Results and Discussion}
We study two rule-based receptor models Fc$\epsilon$RI and T-cell for experimental purposes. We show that the presented method BioMETA is successful in estimating $26$ parameters of a Fc$\epsilon$RI model and $29$ parameters of a T-cell model satisfying three STL properties each. The next two subsections describe the models, discuss their respective STL specifications and demonstrate our experimental results. All experiments were performed on an Intel Xeon Platinum 8160 48-Core 2.10 GHz processor with 768 GB of RAM operating under Ubuntu 16.04.6 LTS.

\subsection{Benchmark 1: Fc$\epsilon$RI}
Fc$\epsilon$RI is the high affinity receptor for immunoglobulin E (IgE) and is a member of the multichain immune recognition receptors (MIRR) family. It is responsible for controlling the activation of two different types of white blood cells known as human mast cells and basophils which are a crucial part of the immune and neuroimmune system in living organisms. To be more specific, Fc$\epsilon$RI plays an important role in wound healing and immediate allergic reactions. It also participates in antigen presentation of immediate hypersensitivity reactions involving IgE-mediated release of histamine and some other mediators. Some of these allergic reactions can result in serious consequences and Fc$\epsilon$RI is of vital importance \cite{turner1999signalling} in providing physiological protection against such reactions and maintaining the allergic response by controlling the secretion of allergic mediators and induction of cytokine gene transcription.

We perform our experiments on a rule-based model of Fc$\epsilon$RI developed by Faeder et al. \cite{faeder2003investigation} which aims to examine the function of multiple components in the phosphorylation and activation of Spleen tyrosine kinase, Syk, whose inhibition aids in treating autoimmune diseases. The authors of \cite{faeder2003investigation} model Fc$\epsilon$RI in its tetrameric form, which includes an $\alpha$-chain that binds IgE and three Immunoreceptor Tyrosine-based Activation Motif (ITAM)-containing subunits, a $\beta$-chain, and two disulfide-linked $\gamma$-chains. The model exhibits four major reactions namely association, dissociation, phosphorylation, and dephosphorylation. All experiments in \cite{faeder2003investigation} are performed to stimulate and observe rat basophilic leukemia (RBL) cells using covalently cross-linked IgE dimers.

We translate three important model behaviors described in the results section of \cite{faeder2003investigation} to the following probabilistic Signal Temporal Logic behavioral properties:

\begin{flushleft}
    \textbf{Property 1:}
    \begin{ceqn}
    \begin{align*}
       P_{\geq 0.95}( \textbf{F}_{[0,1500]}(({RecDim / RecTot} > 0.5)  \land 
            \textbf{G}_{[1500,3000]}({RecDim / RecTot} \geqslant 0.5)) )
    \end{align*}
    \end{ceqn}
\end{flushleft}

The first property investigates the kinetics of receptor aggregation and tyrosine phosphorylation in RBL cells when they are stimulated with covalently linked dimers of IgE. This property monitors one of the features of dimer-induced receptor (RecDim) phosphorylation where the percentage of RecDim reaches half of its maximum value at around half an hour of the simulation and keeps on increasing till one hour. In other words, this property verifies if the percentage of RecDim is observed to persist in the later half of the simulation.

\begin{flushleft}
    \textbf{Property 2:}
    \begin{ceqn}
    \begin{align*}
       P_{\geq 0.80}( \textbf{F}_{[0,1500]}(({LynRecPbeta / LynTot} \geqslant 0.8)  \land         \textbf{G}_{[1500,3000]}({LynRecPbeta / LynTot} \geqslant 0.8)))
    \end{align*}
    \end{ceqn}
\end{flushleft}

Simulating the Fc$\epsilon$RI model shows a rapid redistribution of Tyrosine-protein kinase Lyn. This redistribution becomes even more tightly associated with the receptor via binding of Lyn's SH2 domain to the phosphorylated $\beta$ ITAM (LynRecPbeta). The second property verifies if a large percentage (say 80\%) of the available Lyn is bound through its SH2 domain to $\beta$ when the receptor aggregation reaches its maximum.

\begin{flushleft}
    \textbf{Property 3:}
    \begin{ceqn}
    \begin{align*}
       P_{\geq 0.99}( \textbf{G}_{[1200,3000]}({RecPgamma / RecPbeta} \geqslant 2.0))
    \end{align*}
    \end{ceqn}
\end{flushleft}

\begin{figure}
	\centering
	\includegraphics[width=0.76\columnwidth]{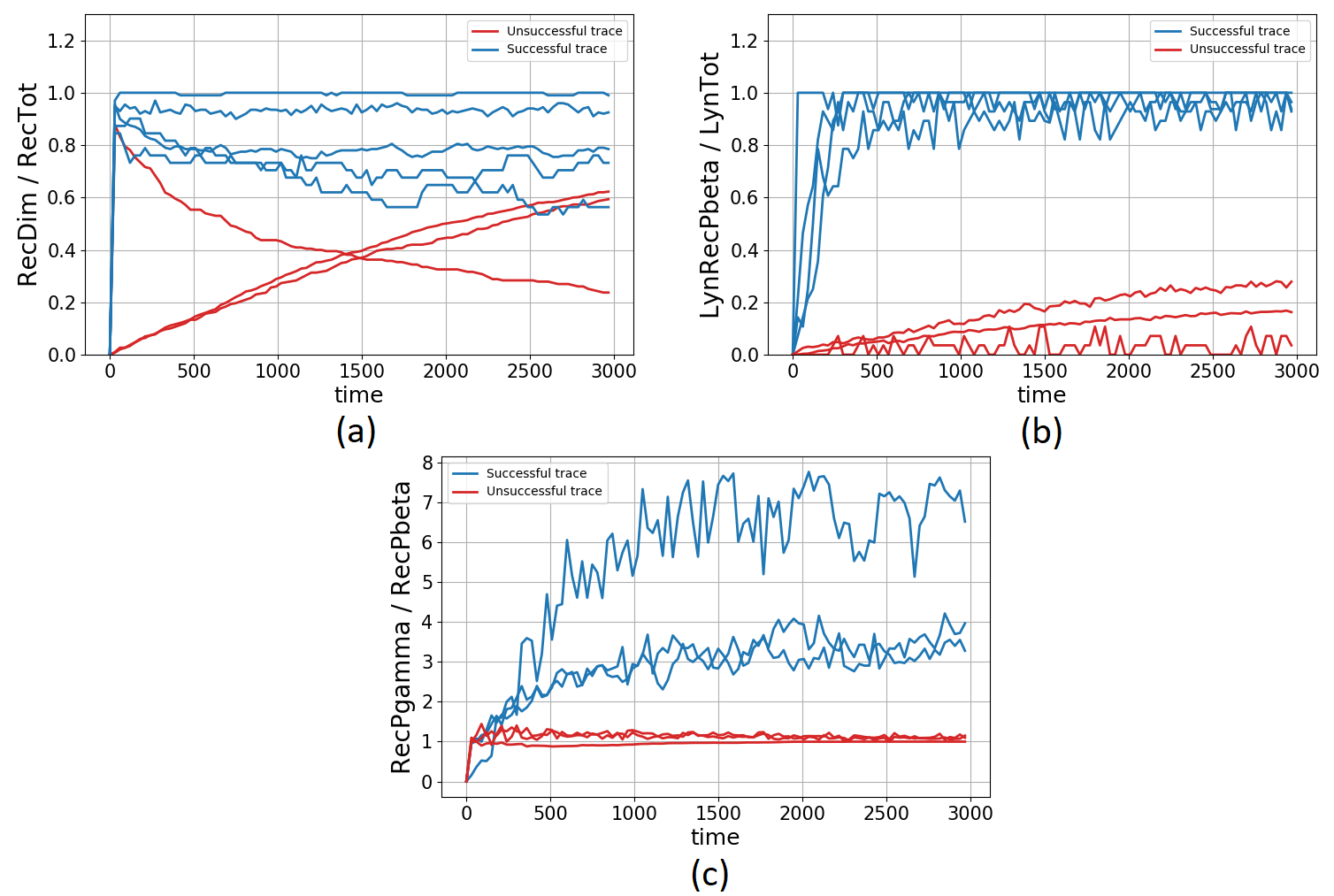}
    \caption{Successful simulation trajectories (in blue) of Fc$\epsilon$RI model satisfying property 1, 2 and 3. Unsuccessful trajectories against each of the above properties are shown in red.}
    \label{figure2}
\end{figure}

Another observation made by the authors in \cite{faeder2003investigation} is that Syk, which binds to the $\gamma$ ITAM, is present in much higher concentration than Lyn, which binds to the $\beta$ ITAM. This happens because a phosphorylated $\gamma$ ITAM tyrosine has a longer lifetime than a $\beta$ phosphotyrosine. Our third property verifies this behavior by showing that the level of $\gamma$ phosphorylation (RecPgamma) exceeds that of $\beta$ phosphorylation (RecPbeta) by $\sim$2-fold after simulating the model for some time and persists this behavior afterwards. Figure \ref{figure2} illustrates the successful traces obtained from BioNetGen verifying the above three STL properties of Fc$\epsilon$RI model. For comparison purposes, this figure also shows some of the unsuccessful traces obtained during the model checking phase.

Table \ref{table_1} shows the estimated set of parameter values of Fc$\epsilon$RI satisfying all three specifications simultaneously obtained by BioMETA when $\alpha = 0.005$, $\beta = 0.2$ and $\pmb{r} = \{0.95,0.80,0.99\}$. As mentioned before there are two types of parameters in rule-based models: initial concentration of species and rate constants. The first four parameters in Table \ref{table_1} represent the molecules present in Fc$\epsilon$RI model and the values represent their initial concentration at the start of each simulation. Parameters 5-26 are the rate constants which measure the rate at which a certain reaction rule proceeds. For instance, parameter 17 (pLb) and 19 (pLg) act as the rate constants for receptor transphosphorylation of $\beta$ and $\gamma$ ITAMs respectively by constitutive Lyn. The result of this transphosphorylation is being observed in Property 3 of Fc$\epsilon$RI model. Each of the rate constants is able to regulate one or more reactions. For example, parameter 25 (dm) serves as a rate constant for two reactions i.e. receptor dephosphorylation of $\beta$ as well as $\gamma$ ITAMs.

\begin{table}
\caption{\normalfont
Estimated set of parameter values obtained by BioMETA for Fc$\epsilon$RI satisfying three specifications with $\pmb{r}$ = \{0.95, 0.80, 0.99\}, $\alpha$ = 0.005, $\beta$ = 0.2}
    \label{table_1}
\centering
\begin{tabular}{lll}
\hlineB{2}
Param No. & Param Name & Param Value \\
\midrule
           1 &Lig\_tot &3.748e06 \\
           2 &Rec\_tot &582.79 \\
           3 &Lyn\_tot &0.5921 \\
           4 &Syk\_tot &1.772e05 \\
           5 &kp1 &1.90e-09 \\
           6 &km1 &1.507 \\
           7 &kp2 &4.1474 \\
           8 &km2 &0.0037 \\
           9 &kpL &0.2603 \\
           10 &kmL &0.5118 \\
           11 &kpLs &0.3815 \\
           12 &kmLs &0.00036 \\
           13 &kpS &0.02757 \\
           14 &kmS &0.0055 \\
           15 &kpSs &0.5274 \\
           16 &kmSs &0.00013 \\
           17 &pLb &61.432 \\
           18 &pLbs &13124.41 \\
           19 &pLg &397.217 \\
           20 &pLgs &0.1716 \\
           21 &pLS &0.1181 \\
           22 &pLSs &1453.91 \\
           23 &pSS &10578.92 \\
           24 &pSSs &5.3064 \\
           25 &dm &0.06354 \\
           26 &dc &73.245 \\
\hlineB{2}
\end{tabular}
\end{table}

\subsection{Benchmark 2: T-cell}
Our second benchmark is a rule-based model of T-cell receptor, also known as T-lymphocyte. It is a type of white blood cell that is a vital part of the immune system. It detects the presence of toxins or other foreign substances, known as antigens, with the help of T-cell receptors (TCRs). The antigens trigger a living organism's immune system to produce anitbodies which in turn help protect the body against bacteria and viruses. The surface receptors TCRs bind to some specific polypeptide fragments that are displayed, by a protein called the major histocompatibility complex (MHC), on the surface of neighboring cells. T-cell is known to maintain a very intricate balance between exhibiting strong responses to the presence of extremely small quantities of antigen while not responding to the large quantities of host peptide-MHC (pMHC). By detecting antigens in this way, it is able to successfully avoid autoimmunity. This characteristic along with some other attributes affiliated with the model, add an element of uncertainty to its dynamics resulting in trajectories that may exhibit completely different behavior even though generated from the same initial state \cite{clarke2008statistical}. Such a stochastic nature of this model makes it a challenging benchmark for our current study.

In order to avoid autoimmunity, T-cell receptor must ignore self peptides and at the same it must recognize foreign peptides for immune defense. Thus, a correct T-cell model is expected to discriminate between agonist and antagonist peptides \cite{lipniacki2008stochastic}. In order to test this particular behavior, we observe the model's primary output i.e. the fraction of doubly phosphorylated ERK (ppERK). This fraction (ppERK/totERK) is also taken as a measure of T-cell activation. If $ppERK/totERK < 0.10$, the cell is considered to be inactive and if $ppERK/totERK > 0.50$, the T-cell is considered to be in its active state. All three of our properties closely observe the value of this fraction and determine if the cell successfully achieves an active state after exhibiting an inactive state (and vice versa) within a defined period of time. We estimate the parameter values of the model such that it satisfies the following three properties:

\begin{flushleft}
    \textbf{Property 1:}
    \begin{ceqn}
    \begin{align*}
       P_{\geq 0.85}( \textbf{G}_{[0,300]}({ppERK/totERK} < 0.95))
    \end{align*}
    \end{ceqn}
\end{flushleft}

As mentioned earlier, cell activity is measured by changing the quantity of ppERK. The first property verifies if the fraction of ppERK always stays below a given threshold value (0.95) during the first 300 time units of the simulation.

\begin{flushleft}
    \textbf{Property 2:}
    \begin{ceqn}
    \begin{align*}
       P_{\geq 0.80}( \textbf{F}_{[0,300]}(({ppERK / totERK} < 0.1) \land 
			\textbf{F}_{[300,600]}({ppERK / totERK} > 0.5)))
    \end{align*}
    \end{ceqn}
\end{flushleft}

Model simulations have shown that a small number of agonist peptides are sufficient for cell activation. Our second property verifies this behavior by determining if T-cell is able to achieve the active state in the later half of its simulation provided that it was inactive during the first half.

\begin{flushleft}
    \textbf{Property 3:}
    \begin{ceqn}
    \begin{align*}
       P_{\geq 0.70}( \textbf{F}_{[0,1000]}(({ppERK / totERK} > 0.5) \land 
			\textbf{F}_{[1000,2000]}({ppERK / totERK} < 0.1)))
    \end{align*}
    \end{ceqn}
\end{flushleft}

\begin{figure}
	\centering
	\includegraphics[width=0.76\columnwidth]{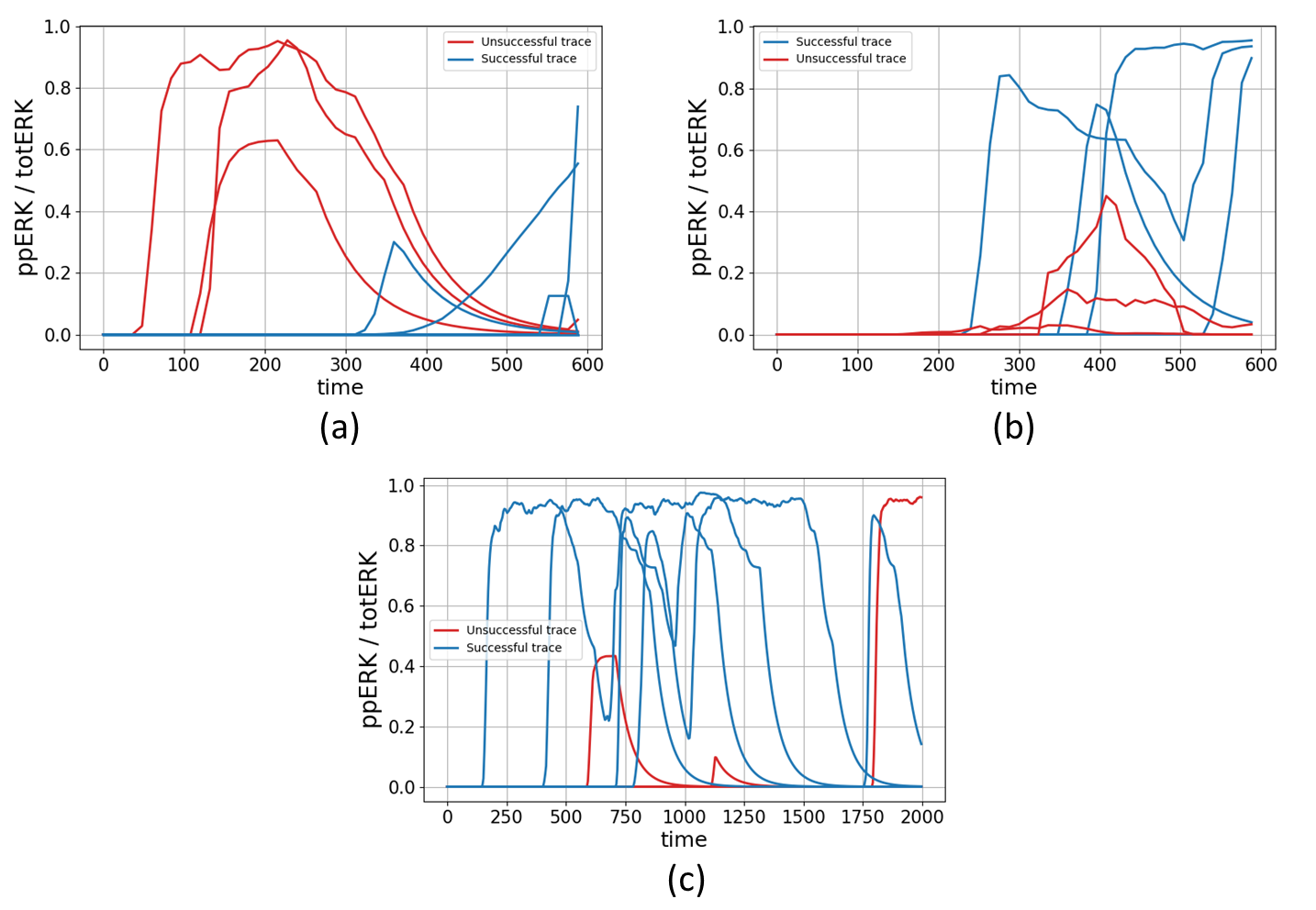}
    \caption{Successful simulation trajectories (in blue) of T-cell model satisfying property 1, 2 and 3. Unsuccessful trajectories against each of the above properties are shown in red.}
    \label{figure3}
\end{figure}

It is known that cell activation achieved by agonist peptides (Property 2) can be almost completely inhibited by antagonist peptides \cite{lipniacki2008stochastic}. The third property monitors the deactivation behavior of T-cells which is a result of pSHP mediated negative feedback. This property verifies if the system can go from an active state during first 300 seconds of the simulation to an inactive state during the next 300 seconds. Figure \ref{figure3} illustrates the successful as well as unsuccessful traces of T-cell model obtained from BioNetGen against the above three STL properties.

Table \ref{table_2} shows the estimated set of parameter values of the same model satisfying all three probabilistic specifications simultaneously obtained by BioMETA when $\alpha = 0.005$, $\beta = 0.2$ and $\pmb{r} = \{0.85, 0.80, 0.70\}$. In Table \ref{table_2}, the first eight parameters represent the molecules present in T-cell model with their respective initial concentrations. Parameters 9-29 are the rate constants which quantify the rate of reaction rules. All three properties validated for this model involve the monitoring of doubly phosphorylated ERK (ppERK). The rate constants involved in the phosphorylation and dephosphorylation of ERK are parameter 28 (e1) and 29 (e2) respectively.



\begin{table}
\caption{\normalfont
Estimated set of parameter values obtained by BioMETA for T-cell model satisfying three specifications with \hspace{5mm} $\pmb{r}$ = \{0.85, 0.80, 0.70\}, $\alpha$ = 0.005, $\beta$ = 0.2}
\centering
    \label{table_2}
\begin{tabular}{lll}
\hlineB{2}
Param No. & Param Name & Param Value \\
\midrule
           1 &pMHC(p$\sim$ag) &2380.744 \\
           2 &pMHC(p$\sim$en) &1.984e05 \\
           3 &TCR &170.82 \\
           4 &LCK &9.002e05 \\
           5 &ZAP &1.121e06 \\
           6 &MEK &7.734e05 \\
           7 &ERK &63578.59 \\
           8 &SHP &2681.37 \\
           9 &b1 &0.000103 \\
           10 &b2 &0.000269 \\
           11 &d1 &0.8554 \\
           12 &d2 &244.886 \\
           13 &lb &8.76e-09 \\
           14 &ly1 &1.90e-06 \\
           15 &ly2 &2.6903 \\
           16 &ls1 &0.0025 \\
           17 &ls2 &1.79e-05 \\
           18 &tp &1.4047 \\
           19 &s0 &3.48e-08 \\
           20 &s1 &0.00107 \\
           21 &s2 &1.22e-06 \\
           22 &s3 &0.000113 \\
           23 &z0 &0.000121 \\
           24 &z1 &0.00168 \\
           25 &z2 &1.52572 \\
           26 &m1 &0.000504 \\
           27 &m2 &0.38002 \\
           28 &e1 &1.32e-05 \\
           29 &e2 &0.5078 \\
\hlineB{2}
\end{tabular}
\end{table}

\section{Conclusions and Future Work}
In this study, we introduce a new approach for estimating the parameters of a rule-based stochastic biochemical model such that the model is able to satisfy multiple probabilistic temporal logic behavioral specifications simultaneously. Our proposed method BioMETA is based on merging a multiple hypothesis testing based statistical model checking approach with a simulated annealing based search to find a single set of parameter values for the model so that the model satisfies all the given probabilistic temporal logic behavioral specifications. Our experimental results demonstrate that BioMETA was successful in estimating $26$ and $29$ parameters of two rule-based biochemical models Fc$\epsilon$RI and T-cell respectively against three probabilistic temporal logic behavioral specifications each. 

We plan to pursue multiple directions for future work. Our current method requires several sequential model simulations to be generated in order to check model satisfiability; therefore, we plan to take advantage of the latest parallel computing frameworks and execute these simulations in parallel by implementing the presented algorithm on GPUs. The curse of dimensionality is one of the biggest challenges faced when designing solutions to such problems. To address this, we intend to investigate dimensionality reduction techniques \cite{jha2017stochastic} which would help the search process to perform more efficiently and generate useful results faster. 

We are also interested in exploring and designing solutions based on deep learning techniques that can improve the search process. Another interesting future direction is to construct a neural network that is able to learn the given biochemical model from its BioNetGen simulation traces. Later this neural network could be used to generate hundreds of simulation traces in parallel on GPUs eliminating the need to actually simulate the model. This could eventually help expedite the parameter estimation process.

\bibliographystyle{unsrt}  
\bibliography{main}

\begin{thebibliography}{10}

\bibitem{faeder2009rule}
James~R Faeder, Michael~L Blinov, and William~S Hlavacek.
\newblock Rule-based modeling of biochemical systems with {BioNetGen}.
\newblock {\em Systems Biology}, pages 113--167, 2009.

\bibitem{blinov2004bionetgen}
Michael~L Blinov, James~R Faeder, Byron Goldstein, and William~S Hlavacek.
\newblock {BioNetGen}: software for rule-based modeling of signal transduction
  based on the interactions of molecular domains.
\newblock {\em Bioinformatics}, 20(17):3289--3291, 2004.

\bibitem{gillespie1977exact}
Daniel~T Gillespie.
\newblock Exact stochastic simulation of coupled chemical reactions.
\newblock {\em The Journal of Physical Chemistry}, 81(25):2340--2361, 1977.

\bibitem{gillespie2009refining}
Dan~T Gillespie, Min Roh, and Linda~R Petzold.
\newblock Refining the weighted stochastic simulation algorithm.
\newblock {\em The Journal of Chemical Physics}, 130(17):174103, 2009.

\bibitem{calzone2006machine}
Laurence Calzone, Nathalie Chabrier-Rivier, Fran{\c{c}}ois Fages, and Sylvain
  Soliman.
\newblock Machine learning biochemical networks from temporal logic properties.
\newblock {\em Lecture Notes in Computer Science}, 4220:68--94, 2006.

\bibitem{hussain2015automated}
Faraz Hussain, Christopher~J Langmead, Qi~Mi, Joyeeta Dutta-Moscato, Yoram
  Vodovotz, and Sumit~K Jha.
\newblock Automated parameter estimation for biological models using {Bayesian}
  statistical model checking.
\newblock {\em BMC Bioinformatics}, 16(17):S8, 2015.

\bibitem{donze2013efficient}
Alexandre Donz{\'e}, Thomas Ferrere, and Oded Maler.
\newblock Efficient robust monitoring for {STL}.
\newblock In {\em International Conference on Computer Aided Verification},
  pages 264--279. Springer, 2013.

\bibitem{jha2017telex}
Susmit Jha, Ashish Tiwari, Sanjit~A Seshia, Tuhin Sahai, and Natarajan Shankar.
\newblock Telex: Passive stl learning using only positive examples.
\newblock In {\em International Conference on Runtime Verification}, pages
  208--224. Springer, 2017.

\bibitem{chylek2014rule}
Lily~A Chylek, Leonard~A Harris, Chang-Shung Tung, James~R Faeder, Carlos~F
  Lopez, and William~S Hlavacek.
\newblock Rule-based modeling: a computational approach for studying
  biomolecular site dynamics in cell signaling systems.
\newblock {\em Wiley Interdisciplinary Reviews: Systems Biology and Medicine},
  6(1):13--36, 2014.

\bibitem{xu2011rulebender}
Wen Xu, Adam~M Smith, James~R Faeder, and G~Elisabeta Marai.
\newblock Rulebender: a visual interface for rule-based modeling.
\newblock {\em Bioinformatics}, 27(12):1721--1722, 2011.

\bibitem{smith2012rulebender}
Adam~M Smith, Wen Xu, Yao Sun, James~R Faeder, and G~Elisabeta Marai.
\newblock Rulebender: integrated modeling, simulation and visualization for
  rule-based intracellular biochemistry.
\newblock {\em BMC Bioinformatics}, 13(8):S3, 2012.

\bibitem{liu2016parameter}
Bing Liu and James~R Faeder.
\newblock Parameter estimation of rule-based models using statistical model
  checking.
\newblock In {\em 2016 IEEE International Conference on Bioinformatics and
  Biomedicine (BIBM)}, pages 1453--1459. IEEE, 2016.

\bibitem{blinovrule}
Michael~L Blinov, James~R Faeder, and William~S Hlavacek.
\newblock Rule-based modeling of biological systems using {BioNetGen} modeling
  language.

\bibitem{raman2015reactive}
Vasumathi Raman, Alexandre Donz{\'e}, Dorsa Sadigh, Richard~M Murray, and
  Sanjit~A Seshia.
\newblock Reactive synthesis from signal temporal logic specifications.
\newblock In {\em Proceedings of the 18th International Conference on Hybrid
  Systems: Computation and Control}, pages 239--248. ACM, 2015.

\bibitem{legay2010statistical}
Axel Legay, Beno{\^\i}t Delahaye, and Saddek Bensalem.
\newblock Statistical model checking: An overview.
\newblock In {\em International Conference on Runtime Verification}, pages
  122--135. Springer, 2010.

\bibitem{zuliani2015statistical}
Paolo Zuliani.
\newblock Statistical model checking for biological applications.
\newblock {\em International Journal on Software Tools for Technology
  Transfer}, 17(4):527--536, 2015.

\bibitem{wald1945sequential}
Abraham Wald.
\newblock Sequential tests of statistical hypotheses.
\newblock {\em The Annals of Mathematical Statistics}, 16(2):117--186, 1945.

\bibitem{palaniappan2013statistical}
Sucheendra~K Palaniappan, Benjamin~M Gyori, Bing Liu, David Hsu, and
  PS~Thiagarajan.
\newblock Statistical model checking based calibration and analysis of
  bio-pathway models.
\newblock In {\em International Conference on Computational Methods in Systems
  Biology}, pages 120--134. Springer, 2013.

\bibitem{hussain2014parameter}
Faraz Hussain, Sumit~K Jha, Susmit Jha, and Christopher~J Langmead.
\newblock Parameter discovery in stochastic biological models using simulated
  annealing and statistical model checking.
\newblock {\em International Journal of Bioinformatics Research and
  Applications 2}, 10(4-5):519--539, 2014.

\bibitem{ramanathan2015parallelized}
R~Ramanathan, Yan Zhang, Jun Zhou, Benjamin~M Gyori, Weng-Fai Wong, and
  PS~Thiagarajan.
\newblock Parallelized parameter estimation of biological pathway models.
\newblock In {\em HSB}, pages 37--57, 2015.

\bibitem{jha2009bayesian}
Sumit~Kumar Jha, Edmund~M Clarke, Christopher~James Langmead, Axel Legay,
  Andr{\'e} Platzer, and Paolo Zuliani.
\newblock A {Bayesian} approach to model checking biological systems.
\newblock In {\em CMSB}, volume 5688, pages 218--234. Springer, 2009.

\bibitem{mancini2015computing}
Toni Mancini, Enrico Tronci, Ivano Salvo, Federico Mari, Annalisa Massini, and
  Igor Melatti.
\newblock Computing biological model parameters by parallel statistical model
  checking.
\newblock In {\em International Conference on Bioinformatics and Biomedical
  Engineering}, pages 542--554. Springer, 2015.

\bibitem{madsen2015biopsy}
Curtis Madsen, Fedor Shmarov, and Paolo Zuliani.
\newblock {BioPSy}: an {SMT}-based tool for guaranteed parameter set synthesis
  of biological models.
\newblock In {\em International Conference on Computational Methods in Systems
  Biology}, pages 182--194. Springer, 2015.

\bibitem{cimatti2002nusmv}
Alessandro Cimatti, Edmund Clarke, Enrico Giunchiglia, Fausto Giunchiglia,
  Marco Pistore, Marco Roveri, Roberto Sebastiani, and Armando Tacchella.
\newblock {NuSMV} 2: An opensource tool for symbolic model checking.
\newblock In {\em International Conference on Computer Aided Verification},
  pages 359--364. Springer, 2002.

\bibitem{wang2016formal}
Qinsi Wang, Natasa Miskov-Zivanov, Bing Liu, James~R Faeder, Michael Lotze, and
  Edmund~M Clarke.
\newblock Formal modeling and analysis of pancreatic cancer microenvironment.
\newblock In {\em International Conference on Computational Methods in Systems
  Biology}, pages 289--305. Springer, 2016.

\bibitem{zhou2017automated}
Jun Zhou, R~Ramanathan, Weng-Fai Wong, and PS~Thiagarajan.
\newblock Automated property synthesis of {ODEs} based bio-pathways models.
\newblock In {\em International Conference on Computational Methods in Systems
  Biology}, pages 265--282. Springer, 2017.

\bibitem{rizk2008continuous}
Aur{\'e}lien Rizk, Gr{\'e}gory Batt, Fran{\c{c}}ois Fages, and Sylvain Soliman.
\newblock On a continuous degree of satisfaction of temporal logic formulae
  with applications to systems biology.
\newblock In {\em International Conference on Computational Methods in Systems
  Biology}, pages 251--268. Springer, 2008.

\bibitem{bartocci2013robustness}
Ezio Bartocci, Luca Bortolussi, Laura Nenzi, and Guido Sanguinetti.
\newblock On the robustness of temporal properties for stochastic models.
\newblock {\em arXiv preprint arXiv:1309.0866}, 2013.

\bibitem{brim2013exploring}
Lubo{\v{s}} Brim, Milan {\v{C}}e{\v{s}}ka, Sven Dra{\v{z}}an, and David
  {\v{S}}afr{\'a}nek.
\newblock Exploring parameter space of stochastic biochemical systems using
  quantitative model checking.
\newblock In {\em Proceedings of the 25th international conference on Computer
  Aided Verification}, pages 107--123. Springer-Verlag, 2013.

\bibitem{vcevska2017precise}
Milan {\v{C}}e{\v{s}}ka, Frits Dannenberg, Nicola Paoletti, Marta Kwiatkowska,
  and Lubo{\v{s}} Brim.
\newblock Precise parameter synthesis for stochastic biochemical systems.
\newblock {\em Acta Informatica}, 54(6):589--623, 2017.

\bibitem{vcevska2016prism}
Milan {\v{C}}e{\v{s}}ka, Petr Pila{\v{r}}, Nicola Paoletti, Lubo{\v{s}} Brim,
  and Marta Kwiatkowska.
\newblock {PRISM-PSY}: precise {GPU}-accelerated parameter synthesis for
  stochastic systems.
\newblock In {\em International Conference on Tools and Algorithms for the
  Construction and Analysis of Systems}, pages 367--384. Springer, 2016.

\bibitem{liepe2014framework}
Juliane Liepe, Paul Kirk, Sarah Filippi, Tina Toni, Chris~P Barnes, and
  Michael~PH Stumpf.
\newblock A framework for parameter estimation and model selection from
  experimental data in systems biology using approximate {Bayesian}
  computation.
\newblock {\em Nature Protocols}, 9(2):439--456, 2014.

\bibitem{bartroff2014sequential}
Jay Bartroff and Jinlin Song.
\newblock Sequential tests of multiple hypotheses controlling type {I} and {II}
  familywise error rates.
\newblock {\em Journal of Statistical Planning and Inference}, 153:100--114,
  2014.

\bibitem{younes2006statistical}
H{\aa}kan~LS Younes and Reid~G Simmons.
\newblock Statistical probabilistic model checking with a focus on time-bounded
  properties.
\newblock {\em Information and Computation}, 204(9):1368--1409, 2006.

\bibitem{turner1999signalling}
Helen Turner and Jean-Pierre Kinet.
\newblock Signalling through the high-affinity {IgE} receptor {Fc$\epsilon$RI}.
\newblock {\em Nature}, 402(6760supp):24, 1999.

\bibitem{faeder2003investigation}
James~R Faeder, William~S Hlavacek, Ilona Reischl, Michael~L Blinov, Henry
  Metzger, Antonio Redondo, Carla Wofsy, and Byron Goldstein.
\newblock Investigation of early events in {Fc$\epsilon$RI}-mediated signaling
  using a detailed mathematical model.
\newblock {\em The Journal of Immunology}, 170(7):3769--3781, 2003.

\bibitem{clarke2008statistical}
Edmund~M Clarke, James~R Faeder, Christopher~J Langmead, Leonard~A Harris,
  Sumit~Kumar Jha, and Axel Legay.
\newblock Statistical model checking in {BioLab}: Applications to the automated
  analysis of {T-cell} receptor signaling pathway.
\newblock In {\em International Conference on Computational Methods in Systems
  Biology}, pages 231--250. Springer, 2008.

\bibitem{lipniacki2008stochastic}
Tomasz Lipniacki, Beata Hat, James~R Faeder, and William~S Hlavacek.
\newblock Stochastic effects and bistability in {T cell} receptor signaling.
\newblock {\em Journal of Theoretical Biology}, 254(1):110--122, 2008.

\bibitem{jha2017stochastic}
Sumit~K Jha and Christopher~J Langmead.
\newblock Stochastic computational model parameter synthesis system, January~31
  2017.
\newblock US Patent 9,558,300.

\end{thebibliography}

\end{document}